\newcommand{\eqr}[1]{(\ref{#1})}
\newcommand{\dagg}{^\dagger}

\documentstyle[prl,aps]{revtex}
\begin{document}
\twocolumn[\hsize\textwidth\columnwidth\hsize\csname@twocolumnfalse%
\endcsname

\draft

\title{Exact realization of $SO(5)$ symmetry
in extended Hubbard models}

\author{Christopher L. Henley}

\address{Dept. of Physics, Cornell University,
Ithaca NY 14853-2501}
\maketitle

\begin{abstract}
\end{abstract}
{Zhang recently conjectured an approximate $SO(5)$ symmetry relating
antiferromagnetic and superconducting states 
in high-$T_c$ cuprates. 
Here, an  exact $SO(5)$ symmetry is implemented in  a
generalized Hubbard model (with long-range interactions) on a lattice.
The possible relation to a more realistic extended Hubbard Hamiltonian
is discussed.}
]

%
%

S.-C.~Zhang recently conjectured that 
high-$T_c$ cuprate compounds possess an
approximate $SO(5)$ symmetry.\cite{zhang-sci}. 
His theory aims to explain the
proximity of superconducting (SC) and antiferromagnetic (AF) phases
in the phase diagram,  and to account for the low-energy excitations 
as approximate $SO(5)$ Goldstone modes.
Antiferromagnetism and superconductivity are unified 
in one grand order parameter field
$(m^x, m^y, m^z, {\rm Re}\Psi, {\rm Im}\Psi)$, behaving
as a 5-component vector, where the first three
elements are Cartesian components of the staggered magnetization and 
$\Psi$ is a spin-singlet SC order parameter 
(Here ``${\rm Re}\Psi$'' $\equiv {1 \over 2} (\Psi + \Psi \dagg)$, etc.)

In this picture, small symmetry-breaking terms tend to 
drive the system in a ``superspin flop'' between 
antiferromagnetism and superconductivity,
just as, 
in an magnet with approximate $SO(3)$ symmetry, 
competing spin-space anisotropies and external field 
can drive a ``spin-flop'' transition between magnetic states
with order along the $z$ axis and in the $xy$ plane.\cite{zhang-sci}

The $SO(5)$ theory, while positing an intimate relationship
between SC and AF order, 
does not imply that the pairing mechanism is 
AF fluctuations~\cite{pines,anderson-nearlyafm}.
Rather, it quantifies the notion
(also relevant to superfluid ${}^3$He)
that there need not be a sharp difference
between interactions mediated by magnetic and ``charge''  (number)
fluctuations.
I pass over Zhang's specific mechanism (whereby 
the system accommodates doping 
by switching from the AF state to a symmetry-related SC state
which has a different particle number), 
for $SO(5)$ symmetry can be valid even 
if another sort of perturbation
is found responsible for the symmetry breaking and the AF-SC transition.

The $41 {\rm meV}$ mode observed in 
spin-flip neutron scattering on {$\rm Y Ba_2 Cu_3 O_7 $}\cite{41mev} is interpreted as
a Goldstone mode  of $SO(5)$ with a gap due to the symmetry-violating
terms, analogous to the anisotropy gap in a spin-wave branch 
of the unixial magnet\cite{zhang-sci}.
These excitations are created by ``$\hat{\pi}\dagg$'' 
operators\cite{demler-prl} ($SO(5)$ generators that mix magnetic and 
SC components).\cite{pi-criticism}
They are charged bosons with the quantum 
numbers of ``preformed'' Cooper pairs, 
and presumably carry the current in the ``normal'' metal\cite{zhang-sci};
it has been speculated\cite{burgess} that 
this explains the linear temperature dependence 
of the normal-state resistivity. 

To the extent that $SO(5)$-violating terms are small
(as in Zhang's phase diagram ``A''\cite{zhang-sci}), 
relations between AF
and SC quantities are obviously predicted. 
For example, the N\'eel temperature $T_N$ on one side ought to
equal the SC $T_c$ on the other side (the real ratio is
$5:1$ in {$\rm Y Ba_2 Cu_3 O_7 $}).
Furthermore, when converted into the proper units, 
the tensors of superfluid density and
AF spin stiffness should be equal;
as should  the order-parameter lengths
(staggered moment  and SC gap magnitude, respectively), and
the interlayer couplings
(interlayer superexchange and intrinsic Josephson coupling, respectively.)
An order-of-magnitude equality of the interlayer couplings
is indeed expected in the interlayer tunneling picture~\cite{chakra-inter}
Finally, the $SO(5)$ Ginzburg-Landau theory predicts that 
vortices have magnetic cores\cite{zhang-sci,arovas}; 
conversely, in analogy to the Bloch wall in the $SO(3)$ 
magnet, it suggests that magnetic domain walls contain 
SC stripes, as proposed for other reasons 
by Emery and Kivelson\cite{stripes}.

{\em Microscopic $SO(5)$ symmetry ---}
In this paper, using elementary notations,
I implement a literal $SO(5)$ symmetry in a one-band lattice model, 
construct a  Hamiltonian with exact $SO(5)$ symmetry, 
and finally consider whether a realistic Hamiltonian of an extended
Hubbard form might approximate an $SO(5)$ symmetric Hamiltonian. 
Take a lattice with $N$ sites (using periodic boundary conditions). 
Creation operators for the orbitals on site ${\bf x}$ are
$c\dagg_\sigma({\bf x})$  for $\sigma = {\uparrow},{\downarrow}$. 
Let ${\bf Q}$ be the ordering wavevector of some two-sublattice 
AF state,
so that $e^{i{\bf Q}\cdot{\bf x}}=\pm 1$ at every site.
The usual staggered-magnetization components are 
 \begin{eqnarray}
   m^{(c)}_z({\bf x}) &\equiv&  {1 \over 2} e^{i{\bf Q}\cdot{\bf x} }
     (c\dagg_{\uparrow}({\bf x})c_{\uparrow}({\bf x}) - c\dagg_{\downarrow}({\bf x}) c_{\downarrow}({\bf x})), \nonumber \\
   m^{(c)}_+({\bf x}) &\equiv& e^{i{\bf Q}\cdot{\bf x} }
     (c\dagg_{\uparrow}({\bf x})c_{\downarrow}({\bf x})) \qquad
  \label{mtagg}
  \end{eqnarray}
and $m^{(c)}_-({\bf x})\equiv m^{(c)}_+({\bf x})\dagg$. 
The SC order parameter operator
has the general form
$\Psi({\bf x}) \equiv \sum _{{\bf r},{\bf r}'}\psi({\bf r}, {\bf r}') c_{\downarrow}({\bf x}+{\bf r}) c_{\uparrow}({\bf x}+{\bf r}')$
(which  allows different spatial symmetries depending on the form of 
the coefficients $\psi({\bf r},{\bf r}')$.) 
We seek a continuous, unitary operation  that turns
a component of ${\bf m}({\bf x})$ into one of $\Psi({\bf x})$, i.e.
turns creation into annihilation operators: 
clearly it must be some form of Bogoliubov transformation. 
 
Indeed, a {\it discrete} $SO(3)$ symmetry of this sort is already
known for the negative-$U$ 
Hubbard model\cite{so3}, 
for which the appropriate SC order parameter is 
$\Psi({\bf x})=c_{\downarrow}({\bf x})c_{\uparrow}({\bf x})$. 
One maps $c_{\downarrow}({\bf x}) \to e^{i{\bf Q}\cdot{\bf x}} c_{\downarrow}({\bf x})\dagg$
(leaving $c_{\uparrow}({\bf x})$ alone) which implies
$(\Psi({\bf x})\dagg, \Psi({\bf x}), e^{i{\bf Q}\cdot {\bf x}} n({\bf x}))$
$ \to (m_+({\bf x}), m_-({\bf x}), m_z({\bf x}))$; here 
$n({\bf x}) \equiv c\dagg_{\uparrow}({\bf x})c_{\uparrow}({\bf x}) + c\dagg_{\downarrow}({\bf x}) c_{\downarrow}({\bf x})$.
The only change induced in the Hubbard Hamiltonian is $U \to -U$;
thus a hidden $SO(3)$ symmetry relates
SC order ($\Psi$) and charge-density-wave order
($e^{i{\bf Q}\cdot{\bf x}} n({\bf x})$) in the limit of large negative $U$.

To write the exact $SO(5)$ symmetry transparently,
and to ensure it in the order parameters and Hamiltonians, 
I use the duality 
\cite{FN-dirac}
between the ``$c$'' operators and 
an alternate set of canonically commuting operators, 
   \begin{equation}
          d_{{\bf k}+{\bf Q},\sigma} \equiv \eta_{\bf k} c_{{\bf k}\sigma}
   \label{dk}
   \end{equation}
In real space, this says
   \begin{equation}
         d_\sigma({\bf x}) = e^{-i{\bf Q}\cdot {\bf x}} \sum _{\bf r} \varphi ({\bf r}) c_\sigma({\bf x}+{\bf r})
   \label{cd-real}
   \end{equation}
where $\eta_{\bf k} \equiv \sum _{\bf r} e^{-i{\bf k}\cdot{\bf r}} \varphi({\bf r})$. 
To make the symmetry \eqr{cprime} work, we will need the important 
conditions\cite{FN-not4n} 
    \begin{mathletters}
    \label{eta-conds}
    \begin{eqnarray}
       \eta_{{\bf k}+{\bf Q}} & = & - \eta_{\bf k} 
       \label{etaeta} \\
       \eta_{-{\bf k}} & = & \eta_{\bf k}
        \label{etasymm}
    \end{eqnarray}
    \end{mathletters}
for all ${\bf k}$, 
which in real space say respectively 
that $\varphi({\bf r})=0$ for ``even'' ${\bf r}$ (meaning those connecting sites in
the same sublattice) and that $\varphi({\bf r})=\varphi(-{\bf r})$.
Eq.~({\ref{etaeta}) implies
        $\{ c\dagg_\sigma({\bf x}), d_{\sigma'}({\bf x}') \} =0$
if ${\bf x}$ and ${\bf x}'$ are on the same sublattice (e.g. ${\bf x}={\bf x}'$). 
\cite{FN-basis}
Then the symmetry operation is just
    \begin{mathletters}
    \begin{eqnarray}
        c'_\sigma({\bf x}) &=& 
            \cos(\phi/2) c_\sigma({\bf x}) 
            + \sin (\phi/2) d\dagg_{-\sigma}({\bf x})
        \\
        d'_\sigma({\bf x}) &=& 
             - \sin (\phi/2) c\dagg_{-\sigma}({\bf x})
             +  \cos(\phi/2) d_\sigma({\bf x}) 
    \label{cprime}
    \end{eqnarray}
    \end{mathletters}
The symmetry \eqr{cprime}
is generated by ${1 \over 2} ({\hat{\pi}}+{\hat{\pi}}\dagg)$, 
where ${\hat{\pi}} = i \sum _{\bf x} [c_{\downarrow}({\bf x}) d_{\uparrow}({\bf x})- d_{\downarrow}({\bf x}) c_{\uparrow}({\bf x})]$. 
To transform wavefunctions, it is useful to know that
the vacuum $|0\rangle$ tranforms to 
$\prod_{\bf k} (\cos (\phi/2) + \eta_{\bf k} \sin(\phi/2) c\dagg_{{\bf k}{\uparrow}}
c\dagg_{-{\bf k}+{\bf Q},{\downarrow}}) |0\rangle $.

So that it will map exactly under \eqr{cprime}, 
the $SO(5)$ staggered magnetization must be defined as
  \begin{equation}
      {\bf m}({\bf x})= {1 \over 2} [{\bf m}^{(c)}({\bf x})
      - {\bf m}^{(d)}({\bf x})]
  \label{mzR}
  \end{equation}
where ${\bf m}^{(d)}$ is \eqr{mtagg} with ``$c$''$\to$``$d$''.
Note this gives a sensible result for the N\'eel state:
if ${\bf m}^{(c)}({\bf x})$ is up, then 
${\bf m}^{(d)}({\bf x})$ is down  (since the $d$ ``orbital'' on
site ${\bf x}$ is a linear combination of ``$c$'' orbitals from the opposite
sublattice).
The SC order parameter is
    \begin{equation}
        \Psi({\bf x})  \equiv  e^{i{\bf Q}\cdot{\bf x}} {1 \over 2} 
               [c_{\downarrow} ({\bf x}) d_{\uparrow}({\bf x}) + d_{\downarrow}({\bf x}) c_{\uparrow}({\bf x})]
    \label{PsiR}
    \end{equation}
Then 
    \begin{mathletters}
    \begin{eqnarray}
           {m}_z{}'({\bf x}) &=& \cos \phi ~m_z({\bf x}) + \sin \phi ~{\rm Re}\Psi({\bf x}) \\
           {\rm Re} \Psi'({\bf x}) &=& \cos \phi ~{\rm Re}\Psi({\bf x}) - \sin \phi ~m_z({\bf x})
    \label{symm}
    \end{eqnarray}
    \end{mathletters}
while the other three components are invariant. 
The $SO(5)$ rotation of the N\'eel state with $\phi=\pi/2$ gives
  \begin{equation}
   2^{-N/2} \prod _{\bf k} (1+\eta_{\bf k} c\dagg_{{\bf k}{\uparrow}} c\dagg_{-{\bf k}{\downarrow}})|0\rangle
  \label{BCS}
  \end{equation}
This BCS state has no remnant of Fermi surface
($\langle c\dagg_\sigma c_\sigma \rangle \equiv  1/2$ throughout reciprocal space.)

One could construct a total of six such rotations, each of which 
mixes one of the three components of ${\bf m}({\bf x})$ with one 
of the two components of $\Psi({\bf x})$.
The other five could all be obtained by 
by combining \eqr{cprime} with the usual $SO(3)$ rotations acting on 
the spin labels of $c$ and $d$ operators,  
plus the usual $SO(2)\equiv U(1)$ gauge symmetry
changing their complex phases. 
(Zhang has discussed the algebra of $SO(5)$ generators\cite{zhang-sci,rabello}.)

For the square lattice, we must have ${\bf Q}=(\pi,\pi)$ 
so the hopping term \eqr{hopk} will be $SO(5)$-invariant.
This leaves much freedom to $\eta_{\bf k}$, 
but the simplest choice is
 \begin{equation}
         \eta_{\bf k} \equiv {\rm sign} (\cos k_x -\cos k_y)
   \label{eta-d}
   \end{equation}
This is inspired by the original and approximate  $SO(5)$
symmetry\cite{zhang-sci} 
which had the same form but with coefficients
$\eta_{\bf k} \to \cos k_x-\cos k_y$; recently
Kohno \cite{rabello} 
independently discovered the exact version \eqr{eta-d}. 
Comparison with \eqr{BCS} shows that \eqr{eta-d}
is essentially the Cooper pair wavefunction and
has $d_{x^2-y^2}$ pairing symmetry,
consistent with strong experimental evidence in the cuprates
\cite{scalapino}. 
Interestingly, one other simple form 
would also satisfy the conditions \eqr{eta-conds}:
$\eta_{\bf k} \equiv {\rm sign} (\cos k_x +\cos k_y)$.
That variant of $SO(5)$, which entails
``extended $s$-wave'' pairing, appears free
from internal contradictions (contrary to a suggestion
in Ref.~\onlinecite{zhang-sci}). 

The coefficients in \eqr{cd-real}
(Fourier transform of \eqr{eta-d}) are
        $\varphi(x,y) = {4/[{\pi^2(x^2-y^2)}} ]$
for $x+y$ odd,  zero for $x+y$ even.
For numerical and analytic explorations, it may also be helpful
to have a one-dimensional toy realization of $SO(5)$ symmetry.
This is given by $Q=\pi$ and
$\eta_{k} \equiv {\rm sign} (\cos k)$, which gives 
$\varphi(r) = 2(-1)^{(r-1)/2}/(\pi r)$ for $r$ odd, zero for $r$ even.

{\em Microscopic Hamiltonian ---} 
Next I will produce an artificial generalization of
the Hubbard Hamiltonian which has exact $SO(5)$ symmetry.
The basic Hubbard model with particle/hole symmetry can be written 
   \begin{eqnarray}
        {\cal H}_{Hubb}= {\cal H}_{hop} + U \sum _{\bf x} {1 \over 2} (n({\bf x})-1)^2,
   \label{Hubb}\\
         {\cal H}_{hop} = (-t) \sum_{{\bf x}\sigma} \sum _{\bf u}
           c\dagg_\sigma({\bf x})  c_\sigma({\bf x}+{\bf u})
         = \sum_{{\bf k}\sigma} 
                   \epsilon_{\bf k} c\dagg_{{\bf k}\sigma} c_{{\bf k}\sigma}
   \label{hopk}
   \end{eqnarray}
with ${\bf u}$ running over nearest neighbors, and
$\epsilon_{\bf k} = (-t) (\cos k_x + \cos k_y)$. 

The minimal Hamiltonian including the terms
in \eqr{Hubb} is simply the $SO(5)$
symmetrization of \eqr{Hubb}. 
The hopping term ${\cal H}_{hop}$ is already invariant 
under all the $SO(5)$ rotations (such as \eqr{symm}), 
{\it provided} that $\epsilon_{{\bf k}+{\bf Q}}=-\epsilon_{{\bf k}}$. 
That is true in any bipartite lattice,  if and only if 
${\bf Q}$ describes the original N\'eel state 
with opposite spins orientations on nearest-neighbor sites.

However, 
$SO(5)$ symmetrization turns the number operator $n({\bf x})$
to something quite different, 
    $n^{s}({\bf x}) \equiv {1 \over 2} [n({\bf x})-n^{(d)}({\bf x})].$
(I take the obvious definition 
$n^{(d)} ({\bf x}) \equiv  d\dagg_{\downarrow}({\bf x}) d_{\downarrow}({\bf x}) + 
                             d\dagg_{\uparrow}({\bf x}) d_{\uparrow}({\bf x})$.)
In contrast to the usual number operator, 
$\sum_{\bf x} n^s({\bf x}) \equiv 0$. 
The $n^{(d)}({\bf x})$ operator includes terms
$|\varphi({\bf r})|^2 n({\bf x}+{\bf r})$, all on the opposite sublattice
from ${\bf x}$ and
largest for nearest neighbors, $|{\bf r}|=1$, as well as long range
hopping between sites of the same sublattice. 
Thus the $SO(5)$-symmetrized Hubbard model has a modified interaction term: 
   \begin{equation}
        {\cal H}^s= {\cal H}_{hop} + U^s \sum _{\bf x} (n^s({\bf x}))^2
   \label{Hubb-so5}
   \end{equation}
When we expand $n^s({\bf x})^2$,
we get a variety of terms, which include interactions and hoppings
(with diminishing coefficients) to arbitrarily large distances. 

What is the ground state of \eqr{Hubb-so5}?
If $U^s\to 0$, at half-filling, it is the Fermi sea which 
manifestly possesses $SO(5)$ symmetry. 
The $t\to 0$ limit of \eqr{Hubb-so5} is more relevant and more challenging.
Certainly a ground state is obtained at half filling
by setting the even sublattice
ferromagnetic in one direction and the odd sublattice ferromagnetic in
any another direction, since $n({\bf x})=n^{(d)}({\bf x})= 1$ on every site. 
The two sublattice moments can be added to make a total angular momentum
$l$ taking any value $\{0,1,\ldots,N \}$. 
As was pointed out by Ref.~\onlinecite{rabello}, 
by applying the ${\hat{\pi}}$ and ${\hat{\pi}}\dagg$ operators, as well as
familiar spin-space rotations, each angular momentum is part of
an $SO(5)$ multiplet with a total degeneracy\cite{eder} 
$(l+1)(l+2)(2l+3)/6$.
This includes states with particle numbers  differing from
$N$ by multiples of $\pm 2$.
The total degeneracy of this family of states is thus
$(N+1)(N+2)^2(N+3)/12$, 
small compared to 
$2^{N}$ in the ordinary Hubbard model \eqr{Hubb} with $t=0$;
however, conceivably this family does not exhaust the ground states.

Now consider how a small $t$ value splits these states in second-order
perturbation theory. Among the states in which each sublattice is 
ferromagnetic aligned, this will have exactly the same effect as
it does in the Hubbard model. This suggests that 
the N\'eel state in fact approximates one of the ground states 
-- and so must \eqr{BCS}, the $SO(5)$ rotation of the N\'eel state, 
since ${\cal H}$ has $SO(5)$ symmetry:
thus I conjecture the $t\ll U^s$ ground state 
{\em has $SO(5)$ broken symmetry.} 

Group theory could be used to enumerate additional allowed terms
in the Hamiltonian as in \onlinecite{rabello}; in particular,
a bilinear coupling of the order parameter on neighboring sites
($SO(5)$ symmetrization of the exchange interaction).
However, I have avoided this $SO(5)$ $t$-$J$ model analog.
It could be derived (in the fashion I just outlined, 
with $|J|\sim t^2/U^s$) from the $SO(5)$ Hubbard-model analog in 
the $t\ll U^s$ limit. 
But the most interesting phases of the standard $t$-$J$ model 
occur in the regime of large $J/t$, which cannot be derived
from any regime of the Hubbard model~\cite{anderson-nearlyafm}, 
and the same thing may happen for the $SO(5)$-Hubbard model \eqr{Hubb-so5}.

{\em Comparison to an extended Hubbard model ---}
I now discuss how one might search for approximate $SO(5)$ symmetry in 
some Hubbard-like model,  such as
   \begin{equation}
       {\cal H}_{ext} = {\cal H}_{Hubb} + {\cal H}'_{hop} + 
        {1 \over 2} V\sum _{{\bf x}{\bf u}} n({\bf x})n({\bf x}+{\bf u})
   \label{Hubb-ext}
  \end{equation}
In \eqr{Hubb-ext} 
${\cal H}'_{hop}$ has the form of \eqr{hopk}, except that 
the coefficient is $t'$ and the displacement is ${\bf u}'$ 
running over {\it second} neighbors.
The last term in \eqr{Hubb-ext} is a Coulomb repulsion between 
nearest neighbor sites. 
Comparison of measurements and  calculations of electronic
structure\cite{H-t2} suggest that $U/t \geq 4$ and
perhaps $t'/t\approx -0.3$ in cuprates. 
The aim is to find
the point(s) in the parameter space of 
${\cal H}_{ext}$ which make it
closest to \eqr{Hubb-so5}:
can the parameters $U$, $V$, and $t'$ of 
\eqr{Hubb-ext} 
be related to $U^s$ in \eqr{Hubb-so5}?

We can very crudely guess at the $U$ 
terms simply by retaining only the terms from \eqr{Hubb-so5}
of exactly this form. They come not only from $n({\bf x})^2$ but also from 
expanding of $n^{(d)}({\bf x})^2$. The result 
is $U= {1\over 4} U^s (1+ \sum _{\bf r} |\varphi({\bf r})|^4)$
$={1\over 4} U^s (1 + 1/9)$. 

We can estimate $V$ in the same fashion, from
the nearest-neighbor term in $n({\bf x})n^{(d)}({\bf x})$
obtaining $V = - U^s |\varphi({\bf u})|^2$ 
where ${\bf u}$ is a nearest neighbor. 
Here the $SO(5)$ symmetry demands an {\it attractive} 
nearest-neighbor electron interaction, which is understandable:
in the $U^s\to\infty$ limit, 
only singly-occupied states can occur in a ground state, so
only the AF states could be ground states.
The SC state
has a certain density of doubly-occupied  and vacant sites,
so an additional term is needed to equalize its energy 
with that of the AF state.
Any pairing interaction might play the same role. 

Finally, no $t'$ terms appear in \eqr{Hubb-so5}. 
In fact, any single-electron hopping term within the same sublattice
violates $SO(5)$ symmetry and gets annihilated in the $SO(5)$ symmetrization.
Thus, although there are many {\it quartic} terms in
\eqr{Hubb-so5} that do hop electrons between sites on 
the same sublattice, there is no $SO(5)$ symmetric way of decoupling
these terms to generate ${\cal H}_{hop}'$. 
(But in second-order 
perturbation theory, those quartic terms can generate e.g. second-neighbor
exchange interactions, just as the $t'$ terms can.)

Very recently, an extended Hubbard model has been diagonalized
using an new interaction with double hoppings,\cite{assaad}
$\sum_{\bf x} K({\bf x})^2$, where 
$K({\bf x}) = 
\sum_{\sigma{\bf u}} c\dagg_\sigma({\bf x}) 
 c_\sigma({\bf x}+{\bf u}) + {c.c.}$. 
That is closer to \eqr{Hubb-so5},  
since its terms have the same form as the largest terms 
(after those already mentioned) in 
$n({\bf x}) n^d({\bf x})$ and $(n^d({\bf x}))^2$. 
This model has an apparently continuous AF/SC transition,~\cite{assaad}
so it may well realize $SO(5)$ approximately.

Of course, even at the $SO(5)$ multicritical point in Zhang's 
picture, the {\it microscopic} Hamiltonian might have no visible 
$SO(5)$ symmetry; 
just as at the ``spin-flop'' point of an anisotropic magnet, 
a cancellation of
terms favoring competing kinds of order might suffice, 
with the symmetry emerging only at long-wavelengths~\cite{zhang-sci}.
But if that length scale is much larger than the numerically 
tractable system size for Hubbard models, 
then direct numerical calculations on finite lattices 
(such as \cite{meixner}) are too small
to address the order parameter symmetry.

Exact diagonalizations (e.g. \cite{meixner})
commonly study ground-state correlations, 
but their spatial decay is often inconclusive
as a test of order due to small system size. 
Yet it is possible that the (excited) eigenstates show a well-defined 
structure characteristic of a particular symmetry;
this provided the convincing evidence 
for long-range order in the spin-$1/2$ triangular lattice 
AF~\cite{bernu-tri}.  
(Very recently, 
Ref.~\cite{eder} has pursued exactly such a program in exact diagonalizations 
of the $t$-$J$ model).
I suggest identifying an $SO(5)$ multiplet numerically in 
a model with manifest $SO(5)$ symmetry,~\cite{eder} 
and then following its evolution while the 
Hamiltonian is adiabatically modified to a more realistic model 
such as \eqr{Hubb-ext}.

{\em Conclusion ---}
I have identified inklings of $SO(5)$ symmetry in popular existing
models and exhibited the form an exact $SO(5)$ symmetry could
take in one- or two- dimensional lattice models.
The $SO(5)$ symmetry in microscopic models
is promising as a spur to the comparison or unification
of competing models of high-$T_c$ superconductivity, 
and to improved understanding of extended Hubbard models. 
However, I have not addressed the murkier 
issue of its application to the cuprates.

Of the objections mounted so far 
to a possible $SO(5)$ relationship between the actual AF and SC phases, 
one seems to be really inescapable: 
the Fermi surface~\cite{baskaran-comment}. 
If the SC metal shows a sharp
drop in electron occupation along a certain surface in 
reciprocal space, as
found in angle-resolved photoemission experiments~\cite{fermisurface}, 
then (see \eqr{dk}) its AF image under $SO(5)$ has a similar surface
(shifted by ${\bf Q}$). 
Apparently this AFM must be a spin-density-wave metal~\cite{FN-fermiliq}.
But the real AF phase of the cuprates is instead deemed to be 
a Mott insulator~\cite{anderson-nearlyafm}, 
and its AF correlations are well modeled using a nearest neighbor 
exchange Hamiltonian\cite{chakra-afm}.

I thank N.-G.~Zhang for calling Ref.~\onlinecite{zhang-sci} to my
attention, and S.-C.~Zhang and S.~Chakravarty
for discussions.
This work was supported by NSF grant 
DMR-9612304.


\begin{references}
\bibitem{FN-basis}
Thus, if we desire, a complete basis for the states is defined using
both $c$ and $d$ operators on (say) the even sublattice, the 
$d$ operators on even sites being related to the $c$ operators on
odd sites by a unitary matrix.
\end{references}
\end{document}